\documentclass[12pt]{article}
\pdfoutput=1
\usepackage{jheppub}
\usepackage{subcaption}
\usepackage{amsmath}
\usepackage{bigints}

\newcommand{\cd}{{\cal D}}

\newcommand{\cm}{{\cal M}}
\newcommand{\cl}{{\cal L}}
\newcommand{\cf}{{\cal F}}

\newcommand{\cj}{{\cal J}}

\newcommand{\ch}{{\cal H}}

\newcommand{\co}{{\cal O}}
\newcommand{\cp}{{\cal P}}

\def\bal#1\eal{\begin{align}#1\end{align}}
\def\alp[#1]{\begin{align}#1\end{align}}

\def\secnum[#1]{\texorpdfstring{$#1$}{TEXT}}

\def\secnuml#1\secnumr{\texorpdfstring{$#1$}{TEXT}}

\def\eqa{\begin{eqnarray}}
\def\eqae{\end{eqnarray}}
\def\eq{\begin{equation}}
\def\eqe{\end{equation}}
\def\be{\begin{equation}}
\def\ee{\end{equation}}
\def\bea{\begin{eqnarray}}
\def\eea{\end{eqnarray}}
\def\ba{\begin{array}}
\def\ea{\end{array}}
\def\bd{\begin{displaymath}}
\def\ed{\end{displaymath}}

\def\Tr{{\rm Tr}}

\def\>{\rangle}
\def\<{\langle}
\def\a{\alpha}
\def\b{\beta}

\def\e{\epsilon}
\def\f{\phi}

\def\l{\lambda}
\def\m{\mu}
\def\n{\nu}

\def\p{\pi}

\def\r{\rho}

\def\t{\tau}

\def\F{\Phi}

\def\L{\Lambda}

\def\pa{\partial}

\title{Ensemble averages, Poisson processes and Microstates}
\author{ Cheng Peng}
\affiliation{Kavli Institute for Theoretical Sciences (KITS), University of Chinese Academy of Sciences, Beijing 100190, China \\
Center for Quantum Mathematics and Physics (QMAP),
Department of Physics,\\ University of California,
Davis, CA 95616 USA}
\emailAdd{pengcheng@ucas.ac.cn  }

\abstract{We consider ensemble averaged theories with discrete random variables. We propose a suitable measure to do the ensemble average. 
	We also provide a mathematical description of such ensemble averages of theories in terms of Poisson point processes. 	Moreover, we demonstrate that averaging theories of this type has an equivalent description as tracing over parts of the microscopic degrees of freedom in a suitable continuous limit of a single microscopic theory.  
	The results from both approaches can be identified with Liouville gravity, 
	of which we further address some implications on the microscopic theory, including venues to look for quantum effects from the view point of the averaged theory. Generalizations to other point processes are also discussed.}

\begin{document}

\maketitle

\section{Introduction}

An increasing amount of evidence  emerges in recent studies that suggests the holographic dual of classical gravity might be the  average of an ensemble of field theories~\cite{Maldacena:2016hyu,Maldacena:2016upp,Maldacena:2017axo,Cotler2017,Saad:2018bqo,Saad:2019lba,Stanford:2019vob,Penington:2019kki,Witten:2020wvy,Afkhami-Jeddi:2020ezh,Maloney:2020nni,Belin:2020hea,Cotler:2020ugk,Maxfield:2020ale,Blommaert:2020seb,Bousso:2020kmy,Stanford:2020wkf,Johnson:2020mwi,Marolf:2020rpm}. 
Most of the analyses by far focus on models with Gaussian type continuous random variables. On the other hand, there has been explicit computations of gravitational path integral in some simple toy models whose boundary dual is shown to be an average of theories where  quantities are subjected to  discrete distributions~\cite{Marolf:2020xie}, see also~\cite{Anous:2020lka,Balasubramanian:2020jhl} for related discussion.  Given their possible connection to the microscopic discreteness of quantum theories,  averages 
measured by discrete distributions are clearly interesting and worth in-depth study on its own right. 

With these motivations, we study properties of an average of field theories with 
random variables  
drawn from  
the Poisson distribution.
We show that the effective theory after the average  
is a Liouville theory.
Along the derivation, we demonstrate the importance of choosing an appropriate measure for the discrete averaging process to give a well behaved effective theory. 
To provide a mathematically more accurate description of such averaging, we show that our setting can be cast precisely into a point process. In this language, the averaging to get the effective action is nothing but the Laplace  functional of the Poisson process.

Moreover it is interesting to understand the nature of averaging over random theories~\cite{Saad:2019lba,Penington:2019kki,Almheiri:2019psf,Almheiri:2019qdq,McNamara:2020uza,Pollack:2020gfa,Langhoff:2020jqa}, namely if the average is genuinely among different theories, or it is simply a useful trick for certain computation, or it is originated from averaging among an ensemble of states in a ``Parent" theory.  
We try to understand this question quantitatively in our model and show that one can rewrite the average over the theories with Poisson randomness as a trace over a part of the microscopic degrees of freedom in a single (suitably double scaled) microscopic model. This connection is different from  previous discussions in the literature, 
and gives a concrete realization that sets up an equivalence between the average over an ensemble of theories and the average over an ensemble of states in a given theory.  
In this microscopic point of view, the above requirement of choosing an appropriate measure in the random average approach is reflected on a careful definition of how to trace out part of the underlying degrees of freedom.

We further discuss the average over random theories subjected to the Skellam distribution. The averaged effective theory is a Sinh-Gordon type model. One can also obtain this resulting theory from tracing out some fermionic degrees of freedom in a double-scaling limit of a microscopic theory.
As in the Liouville case, this gives a concrete realization of a gravitational theory as an effective description of some microscopic model after we choose to erase (part of) the model's microscopic information.

\section{Averaged Poisson random models 
}\label{avg}

Motivated by the above derivation, we consider a real scalar field with a chemical potential.
In Euclidean signature, this is given by the Lagrangian 
\bal
\cl(\f) = \pa_\m \f \pa^\mu \f - J \f\ .\label{freeaction}
\eal
Notice that the discussion in this section applies to general dimension so we will first keep the dimension unspecified.  

In the following we consider a general case where the source $J$ has two components, namely
\bal
J=J_0(x) + J_1(x)\ . 
\eal 
The component $J_0(x)$ is a conventional classical source, and the other $J_1(x)$ is a random source related to a Poisson distribution.

What we are interested in is to consider the effective theory after averaging over the random source $J_1(x)$. \footnote{Notice that we consider this model to be a random average of different theories because what we did is to identify a Lagrangian or Hamiltonian with a fixed ``source" function to the $\f$. Given each such a source, the dynamics of the $\f(x)$ field is uniquely determined. Changing the value of the source in general gives a different theory, although the dimension of the Hilbert space do not change for generic values of the source. In other words, since in this model we do not assume the $J_1(x)$ field to be dynamical and it is not included in the path integral, changing it means changing the definition of the theory.} In practice, this means we would like to find $S_{\text{eff}}$ schematically from
\bal
e^{-S_{\text{eff}}}=\int\mathcal{D}{J_1(x)}\cp(J_1(x)) e^{- \int d^d x \cl(\f) }\ .
\eal
The crucial question is how to pick the correct measure $\mathcal{D}{J_1(x)}\mathcal{P}(J_1(x))$. 
One might think this is in exact parallel to the average over a Gaussian type random coupling like in the SYK model~\cite{Sachdev:1992fk,PG,Maldacena:2016hyu,Polchinski2016,Jevicki2016a,Kitaev:2017awl} or models with Gaussian random sources~\cite{PhysRevLett.37.1364,PhysRevLett.43.744,PhysRevLett.46.871,VanRaamsdonk:2020tlr,Kaviraj:2019tbg,Kaviraj:2020pwv}, where $\cp(J_1(x))= $Pois$(J_1(x),\l(x))$ being simply the Poisson distribution with parameter $\l(x)$. However, as we show in appendix~\ref{wrong} this  naive definition does not give a sensible average over the discrete valued random sources. 

In fact, it turns out that it is not at all trivial to pick a correct measure to do the averaging over such a set of theories; an inappropriate choice could leads to pathological (to all degree) resulting theories. 
In the following, we give an example of finding a sensible measure of the set of theories to be averaged. We provide 2 different approaches, one more physically oriented while the other more mathematically rigorous, to analyse this discrete random model and we will show that they lead to the same result.

\subsection{A physical point of view}\label{randphy}

To make the average process well defined, it turns out that we should treat the combination $ J_1(x) dV(x)$ as the random variable that satisfies, in the example of Poisson distribution
\bal
P(J_1(x))=\prod_{n}{\text{Pois}}\left( J_1(x_n) dV(x_n),\l(x_n) dV(x_n) \right)\,, \quad \forall\, dV(x_n) \text{ s.t. } \sum_n dV(x_n)=\cm\,,\label{res1}
\eal 
where $dV(x)$ is an volume element around position $x$ and   $\l(x) dV(x)$ is the Poisson  parameter. In writing this expression we have chosen a given discretization of the spacetime, so that the volume $V(\cm) = \sum dV(x_n)$, and we will consider the fine grained limit of such discretization.  

Notice that the Poisson probability distribution we considered here has a position dependent parameter, and the distribution at each point only depends on the local information. The latter is the same as the assertion that currents on different $dV(x_n)$ are mutually independent for any discretization, which is the reason that we can write it as a product as in~\eqref{res1}. The probability function superficially depends on how we do the discretization of the spacetime $\cm$, but as we will show in the following, since the mutually independence property is true for any discretization,  
all the results after an average over this probability distribution does not depend on the concrete discretization, and hence we can take the continuum limit of the spacetime discretization smoothly.    

It is crucial that the distribution is integer valued, rather than real valued, so that we can treat it as a counting measure supported on a measure zero subset in an arbitrarily small $dV$; the discretized value $J_1(x) dV(x) $ can be thought of as counting the number of random points in the volume $dV(x)$. Therefore putting $dV$ into the distribution is as well defined as an integral over a sum of Dirac delta functions in $dV(x)$. In addition, we have also rescaled the $\l(x_n)$ accordingly so that the distribution itself is not singular, i.e. $\l(x)$ is not zero, as the volume element tends to zero. To put it another way, this can be understood as the following: if we consider the mean of the rescaled sources $ J_1(x_n)dV(x)$ that is subjected to a Poisson distribution with parameter $\l'(x)$
\bal
\<J_1(x) dV(x) \>_{\l'(x)} =\sum_{k} \text{Pois}( J_1dV(x)=k,\l'(x)) ( J_1(x) dV(x))=  \l'(x) \ . 
\eal 
Physically, we would like to appropriately normalize the mean value, which means we would like the mean value of $dV(x) J_1(x_n)$ to again be proportional to the volume element $dV(x)$. Therefore we choose to rescale the mean value to  
\bal
\l'(x)=dV(x) \l(x)\ .
\eal 
This mean value is just the parameter in the Poisson distribution, so the above scaling argument indicates that we have to consider the Poisson distribution with the parameter $\l(x_n)dV(x_n)$. This then justifies why we would consider the rescaled distribution~\eqref{res1}.

With this choice of the random ensemble, we can work out the average of the random sources of the model. We will formulate the computation in more rigorous mathematical language in the next section. Here we provide an instructive derivation where we discretized the integral as a  Riemann sum and then take the continuous limit
\bal
&\<e^{\int dV(x) J_1(x)\f(x)}\>_{J_1}=\sum P(J_1(x) dV(x) ,\l(x) dV(x) )e^{\int dV(x) J_1(x)\f(x)}\label{aveerage}\\
&=\left(\prod_{n}\sum_{  k=0}^{\infty}\text{Pois}\left( J_1(x_n)dV(x_n)=k,dV(x_n)\l(x_n)\right)\right) e^{ \sum_n dV(x_n) J_1(x_n)\f(x_n) }\\
&=\prod_{n}\left(\sum_{k=0}^{\infty}\text{Pois}\left( J_1(x_n)dV(x_n)=k,dV(x_n)\l(x_n)\right) e^{ dV(x_n) J_1(x_n)\f(x_n) }\right)
\\
&=\prod_{n}\left( e^{ dV(x_n)\l(x_n) \left(e^{\f(x_n)}-1\right) }\right)= e^{\sum_n dV(x)\l(x_n) \left(e^{\f(x_n)}-1\right) }\\
&=\exp\left(\int dV(x) \l(x)\left(e^{\f(x)}-1\right)\right)\ .\label{disc2}
\eal
Notice that in the second line we have used the fact that the Probability distribution at different positions are mutually independent. We have also used $n$ to collectively label the different grid points.

This leads to the following averaged effective potential 
\bal
&\int \mathcal{D}{J_1(x)}P(J_1(x)dV(x),\l(x)dV(x)) e^{ \int dV(x) J_1(x)\f(x) }\\
&\qquad\qquad\qquad :=\<e^{\int dV(x) J_1(x)\f(x)}\>_{J_1}=e^{\int dV(x) \l(x)(e^{\f(x)}-1)}\,,
\eal
where the ``path integral" $\mathcal{D}{J_1(x)}$  contains both a sum over all the sources at different spacetime points $x$ and a sum over all possible values of $J_1(x)dV(x)$ at each point $x$. 
Therefore we get the following effective action 
\bal
S_{\text{eff}}= \int d^d x_E \left(\pa_\m \f \pa^\mu \f - J_0(x) \f -\l(x)(e^{\f(x)}-1)\right)\ .\label{Seff}
\eal

The result~\eqref{Seff} is a  generalized Liouville theory, with an effective background curvature indicated by the value of $J_0(x)$ and a ``cosmological constant" $\l(x)$. In this derivation $\l(x)$ remains a function, or a background field, with out dynamics, therefore for any given $\l(x)$ we get a different effective action. Clearly when the Poisson parameter  $\l(x)$ takes a homogeneous value $\l(x)=\l$, the action is exactly a Liouville action. 
We will provide a detailed interpretation of this position dependent expression and its relation to 2D gravity in section~\ref{gravity}. 

Further notice that 
the sign of the Liouville potential in~\eqref{Seff} is ``wrong": from the probability interpretation $\l(x)\geq 0$, on the other hand the potential in Euclidean signature with this sign is unstable. 
To cure this problem, we can define a slightly modified Poisson average procedure. In particular, we can consider the same set of theory with the random potential as in~\eqref{freeaction}. The source still contains a fixed piece $J_0(x)$ and a random piece $J_1(x)$, and the $J_1(x)$ piece is again related to a Poisson distribution. 
The crucial difference is that we do not do the average of this family of theories naively according to the probability distribution. Instead, we consider the average with an extra insertion of the $(-1)^{\cf}$ ``operator" in the measure.  The average of the potential term is  
\bal
\nonumber &\<e^{\int dV(x) J_1(x)\f(x)}\>_{J_1,t}=\sum P(J_1(x) dV(x) ,\l(x) dV(x) )(-1)^{\cf}e^{\int dV(x) J_1(x)\f(x)}\\
&=\left(\prod_{n}\sum_{  k=0}^{\infty}\text{Pois}\left( J_1(x_n)dV(x_n)=k,dV(x_n)\l(x_n)\right)\right) e^{ \sum_n dV(x_n) \left(J_1(x_n)\left(\f(x_n)+i\p \right)+2\l(x)\right) }\\
&=\prod_{n}\left(\sum_{k=0}^{\infty}\text{Pois}\left( J_1(x_n)dV(x_n)=k,dV(x_n)\l(x_n)\right) e^{ dV(x_n) \left( J_1(x_n)\left(\f(x_n)+i\p \right)+2\l(x)\right)  }\right)
\\
&=\prod_{n}\left( e^{ dV(x_n)\l(x_n) \left(e^{\left(\f(x_n)+i\p \right) }+1\right) }\right)= e^{\sum_n dV(x)\l(x_n) \left(e^{\left(\f(x_n)+i\p \right) }+1\right) }\\
&=\exp\left(\int dV(x) \l(x)\left(-e^{\f(x)}+1\right)\right)\,,\label{disc3}
\eal
where 
\bal
(-1)^{\cf}\equiv (-1)^{\int J_1(x) dV(x)}e^{2\int dV(x) \l(x)}\,,\label{twist}
\eal
The $J_1$ independent factor is introduced so that the average is correctly normalized
\bal
\sum P(J_1(x) dV(x) ,\l(x) dV(x) )(-1)^{\cf}\equiv \sum \cp(J_1(x),dV(x))=1\ .
\eal    
Further notice that the inclusion of this twist operator does not affect other properties of the definition. 

With this new twist operator inserted, we obtain the following effective action
\bal
S_{\text{eff}}= \int d^d x_E \left(\pa_\m \f \pa^\mu \f - J_0(x) \f +\l(x)(e^{\f(x)}-1)\right)\ .\label{Seff1}
\eal
where $\l(x)\geq 0$.

Let us summarized what we have done so far. 
Conceptually, the logic we follow here is to consider different ensemble average schemes and check if any of the scheme has a clear physical interpretation. As in many other recent discussions of ensemble averaging of theories, e.g.~\cite{Afkhami-Jeddi:2020ezh,Maloney:2020nni}, right now we do not have a clear criterion to determine what family of theories should be grouped together and averaged over, and what is the measure we should use for the average. In this work, by comparing the result with other well-defined theories we determine what is a better averaging scheme given a set of theories to be averaged over. Therefore an alternative interpretation of the previous analysis is that we find a proper measure for the average of the set of theories~\eqref{freeaction} that necessarily includes a ``twisted" factor~\eqref{twist}. 

In fact, this twist factor resembles the $(-1)^F$ factor in the computation of the Witten index~\cite{Witten:1982df} in supersymmetric theories. The need of this factor  gives a clear indication, together with the fact that the Poisson distribution is discrete, that there must be a microscopic origin of this random averaged model, and in addition there must be fermionic degrees of freedom in the microscopic model. We will see in section~\ref{micro} that this is indeed the case and the Fermionic number $(-1)^F$ operator in the microscopic theory indeed plays a role of the $(-1)^\cf$ operator we inserted here in the measure of the random averaged model.

\subsection{A Poisson process point of view}\label{randmath}

In this section we provide a more mathematical, yet still intuitive, description of the above computation.
The model we are interested in is still 
\bal
\cl(\f) = \pa_\m \f \pa^\mu \f - J \f\,, \qquad J=J_0+J_1\ .
\eal
where now we interpret the source $J_1(x)$ be a Poisson process on the carrier space $X=R^d$. 

\begin{figure}[t]
	\centering
	\includegraphics[width=0.5\linewidth]{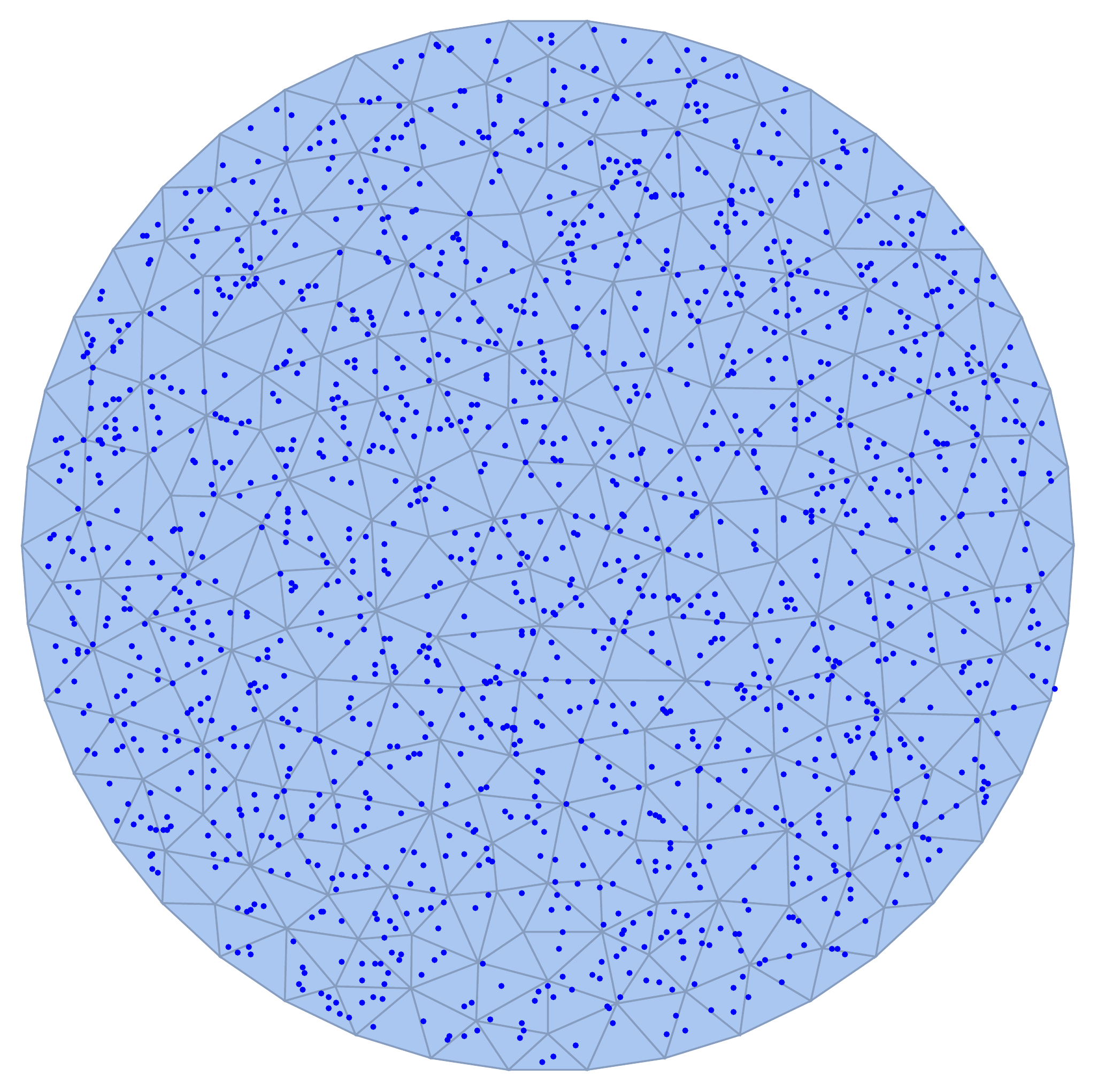}
	\caption{A numerical simulation of a Poisson process. The blue dots represent the Poisson process on a two dimensional carrier space (the grey background). The intensity measure $\l(x) = 1$. The background is a random discretization of the carrier space and each element of the Borel subset can be chosen as a union of cells in this discretization. In this example the intensity measure is a constant which represents a homogeneous Poisson process.}
	\label{fig:pp0}
\end{figure}

There are different equivalent definitions of Poisson distributions. 
Here we adopt the following intuitive definition of the Poisson process~\cite{kingman1992poisson}. A {\bf Poisson process} $\Pi$ describes a random set of points on a given carrier space whose appearing probability is mutually independent and obeys a Poisson distribution. Concretely, let $B$ be a Borel measurable subset of the carrier space $X$. Let the number of the points appearing in this region $B$ be 
\bal
N(B)=\#(\Pi \cap B)\,,\label{nb}
\eal
which defines a counting measure.
Furthermore, for any set of disjoint subsets $B_1,\ldots, B_n$ the Poisson variable $N(B_i)$ are mutually independent. Each $N(B)$ satisfies a Poisson distribution
\bal
P(N(B)=n)=\frac{\L(B)^n}{n!}e^{-\L(B)}\,,
\eal      
where the Poisson parameter $\L(B)$, also known as the mean measure, is determined by 
\bal
\L(B)=\int_B \l(x) dV(x) \,,\label{Ll}
\eal 
where $dV(x)$ is a volume element.   
The integrable function $\l(x)$ is commonly referred to as the intensity function. 
A numerical simulation of a Poisson process is shown in figure \ref{fig:pp0}.

In particular, for infinitesimal $B$ where the integral is well approximated by the area element, we have locally
\bal
\L(d x) = \l(x) dx\ . \label{localL}
\eal
This integrated measure is also the expectation value of the number of points in the above Poisson distribution $\L(B)=\mathbb{E}[N(B)]$. To show this, we consider
\bal
\mathbb{E}[N(B)]&=\sum_n P(N(B)=n) N(B)=\sum_n n \frac{\L(B)^n}{n!}e^{-\L(B)}\\
&=\L(B)\sum_n  \frac{\L(B)^{n-1}}{(n-1)!}e^{-\L(B)}=\L(B)\ .
\eal
The $\L(x)$ and $\l(x)$ functions parametrize the mean value of the Poisson distribution as a function of $x$, effectively this describes the spatial shape of the Poisson distributions on the carrier space. 

Notice that not any $\L(x)$ ($\l(x)$) can be a mean (intensity) measure. As shown in~\cite{kingman1992poisson}, to make sure that a Poisson process exists, the mean measure needs to be non-atomic, which means the $\l(x)$ should not have any delta function support. Furthermore the mean measure $\L(dx)$ should satisfy the following very mild finiteness condition
\bal
\L(\cm) =\sum_{n=1}^{\infty} \L_n(\cm)\,,\qquad \L_n(\cm) <\infty\ . 
\eal
By the restriction theorem~\cite{kingman1992poisson},we can always formally consider a discretization of the carrier space $\cm = \sum_{n=1}^{\infty} dV(x_n) $ \footnote{Here we have abused the notation of $dV(x_n) $ to represent both the open set and its volume. } and further  decomposes the mean measure $\L$ to a sum $\L_n$ each of which only has a support on $dV(x_n)$, which means $\L_n(A)=\L_n(A\cap dV(x_n))$ for $\forall A\subset \cm$. Therefore as long as the $\L_n(\cm)=\L(dV(x_n))<\infty$, 
there is a Poisson process with the given mean measure $\L$ or equivalently the intensity measure $\l(x)$, even if $\Lambda(\cm) \to \infty$.

Now we come back to our model, the source being a Poisson process means the source, accumulated on a given open set of the carrier space, is identified as the Poisson random variable 
\bal
\int_B dV(x) J_1(x) \sim N(B)\ .
\eal 
In other words, we identify the counting measure on a volume element in our model to be
\bal
N(dx) \sim J_1(x)dV(x)\equiv \cj(dx) \ .\label{countm}
\eal 
Then the interaction part of the action can be understood as
\bal
\int dV(x) J_1(x) \f(x) = \int \cj(d x) \f(x)\ . \label{source}
\eal

In this language the average over this random source is nothing but the expectation value of the exponential 
\bal
\mathbb{E}[ e^{\int dV(x) J_1(x) \f(x)}]=\mathbb{E}[ e^{\int \cj(dx) \f(x)}]\ .
 \eal
Given the identification~\eqref{countm}, this is simply the Laplace functional of the Poisson process. For a general Poisson process with the counting measure $N(dx)$ and mean measure $\L(x)$, the Laplace transform of a test function $f(x)$ is
\bal
\mathbb{E}\left[e^{\a \int_{\cm} N(dx) f(x)}\right]&=e^{\int_{\cm}\L(x)(e^{\a f(x)}-1)}\ .
\eal
For completeness, we provide some details of the Laplace transform in appendix~\ref{pps}.  
Using this result, the quantity we would like to compute in our model thus becomes
\bal
\mathbb{E}[ e^{\int dV(x) J(x) \f(x)}]=\mathbb{E}[ e^{\int \cj(dx) \f(x)}]=e^{\int_{\cm}  \L(x)(e^\f(x)-1)}\,,
\eal

As in the previous more physical derivation, there is a sign flip of the Liouville potential term. In this approach the effect of inserting a $(-1)^n$ term can be equivalently performed by considering a slightly modified point process where the distribution is the Poisson distribution with an extra alternating factor:
\bal
P(N(B)=n)=\frac{\left(-\L(B)\right)^n}{n!}e^{\L(B)}\ .\label{p2}
\eal      
Although the extra sign could make the classical probability interpretation of the $P$ function obscure, it is perfectly compatible with the definition of point process, in particular the independence among different spatial regions. So we can simply consider it as a different measure defining a new point process, with which we can compute the expectation value   
\bal
\mathbb{\tilde{E}}[N(B)]&=\sum_n n \frac{\left(-\L(B)\right)^n}{n!}e^{\L(B)}=-\L(B)\sum_n  \frac{\left(-\L(B)\right)^{n-1}}{(n-1)!}e^{\L(B)}=-\L(B)\ .
\eal
Similarly, the Laplace functional that is crucial in the above definition becomes
\bal
\mathbb{\tilde{E}}[ e^{\int d^d x J_1(x) \f(x)}]=\mathbb{\tilde{E}}[ e^{\int \cj(dx) \f(x)}]=e^{-\int_{\cm}  \L(x)(e^\f(x)-1)}\,,
\eal
The rest computation is identical to those in the previous section and we again arrive at the effective Lagrangian~\eqref{Seff1}.

\subsection{Quenched vs annealed: the ``wormhole" contribution}

We can compute the partition function of the averaged theory, as well as the average of the partition function of the individual theories. 
The difference between the two should be related to the  contributions from ``wormholes" connecting different realizations~\cite{Penington:2019kki,Almheiri:2019qdq,Marolf:2020xie,Engelhardt:2020qpv}.

To be precise, we consider $n$ replicas with either quenched or annealed random variable among them. The annealed partition function of the averaged theory with the Lagrangian~\eqref{Seff1} 
is simply the $n^{\text{th}}$ power of the averaged partition function where the random variables fluctuates in each replica
\bal
\overline{Z}^n&=\left(\int \mathcal{D}\f e^{-S_{\text{eff}}}\right)^n=\int \mathcal{D}\f_1\ldots \mathcal{D}\f_n e^{-\sum_{j=1}^n S_{\text{eff}}(\f_j)}\\
&=\int \mathcal{D}\f_1\ldots \mathcal{D}\f_n e^{-\sum_{j=1}^n \left(\int d^d x_E \left(\pa_\m \f_j \pa^\mu \f_j - J_0(x) \f_j +\l(x)(e^{\f_j}-1)\right)\right)}\ .\label{zbarn}
\eal    
The quenched partition function is the average of the partition functions on the $n$ replicas, where the random variables are not averaged over in each replica and is only averaged over for the $n$ replicas as a whole. This can be computed as
\bal
\overline{Z^n}&=\int \mathcal{D}\f_1\ldots \mathcal{D}\f_n \int\mathcal{D}{J_1(x)}P(J_1(x)) e^{- \int d^d x \sum_{j=1}^n\cl(\f_j)}\\
&=\int \mathcal{D}\f_1\ldots \mathcal{D}\f_n \int\mathcal{D}{J_1(x)}P(J_1(x)) e^{- \int d^d x \sum_{j=1}^n \left(\pa_\m \f_j(x) \pa^\mu \f_j(x) - J(x) \f_j(x)\right)}\\
&=\int \mathcal{D}\f_1\ldots \mathcal{D}\f_n  e^{-   \int d^d x \left(\sum_{j=1}^n\pa_\m \f_j(x) \pa^\mu \f_j(x) - J_0(x) \sum_{j=1}^n\f_j(x)+\l(x)(e^{\sum_{j=1}^n \f_j(x)}-1)\right)}\ .\label{znbar}
\eal
It is clear that the two results~\eqref{zbarn} and~\eqref{znbar} are different, which indicates that in the gravitational dual of this model the wormhole solutions connecting the different boundaries should give significant contributions to the gravitational path integral.

To better illustrate the difference between~\eqref{zbarn} and~\eqref{znbar}, we consider a special case where $\l(x)=\l\gg 1$, then the path integrals can be approximated by the contributions from the saddle points. In this limit, we have
\bal
\overline{Z}^n&=\int \mathcal{D}\f_1\ldots \mathcal{D}\f_n e^{-\sum_{j=1}^n \left(\int d^d x_E \left(\pa_\m \f_j \pa^\mu \f_j - J_0(x) \f_j(x) +\l(e^{\f_j(x)}-1)\right)\right)}\\
&\approx\int \mathcal{D}\f_1\ldots \mathcal{D}\f_n \prod_j\delta(\f_j(x))=1\,, 
\eal  
where the saddle point is at $\f_j(x)=0$ for any $j$.
On the other hand, we get
\bal
\overline{Z^n}&=\int \mathcal{D}\f_1\ldots \mathcal{D}\f_n  e^{-   \int d^d x \left(\sum_{j=1}^n\pa_\m \f_j(x) \pa^\mu \f_j(x) - J_0(x) \sum_{j=1}^n\f_j(x)+\l(x)(e^{\sum_{j=1}^n \f_j(x)}-1)\right)}\\
&\approx\int \mathcal{D}\f_1\ldots \mathcal{D}\f_n \delta(\sum_j\f_j(x))=\int \mathcal{D}\f_1\ldots \mathcal{D}\f_{n -1}\,, 
\eal  
which is divergent. From this result, it is clear that the quenched and the annealed partition functions are significantly different, hence confirming the contribution from wormhole type topologies. 
Moreover, there is a rather simple explanation of the divergence in the ``quenched" partition function $\overline{Z^n}$: its divergence is due to the appearance of $n-1$ zero modes in the theory and from the potential gravity dual interpretation they should correspond to $n-1$ free moduli parameters characterizing different topologies connecting  $k=2,3,\ldots,n$ replicated boundaries.      

Further notice that since we are considering the partition function, instead of the extensive quantities such as the free energy or the entanglement entropy, the second replica discussed in details in~\cite{Engelhardt:2020qpv} is not necessary in our discussion.

\section{Poisson random average from tracing over microstates}\label{micro}

In spite of the recent progresses that demonstrate the success and power of ensemble averaging of theories,
A general subtlety caused by considering an ensemble average of theories is its tension with the traditional point of view of quantum theories. The quantization is usually carried out for a given theory with a single fixed action, which could be an obstruction to further understand ensemble averaged theories and in particular its quantum counterpart. 
A way out is to consider the ensemble average of theories as and effective description of the low energy limit of (a subsystem of) a microscopic theory. In this section, we materialise this idea into an explicit connection that reformulate the above average over the Poisson random potentials  
into a trace over a large number of microscopic degrees of freedom in a single refined model.

\subsection{The microscopic setting}

We consider the following microscopic model. 
The model is defined on a  spatial lattice on each site of which resides a $d$-level spin system. The lattice points are labelled by a ``position" vector $x$. 
\footnote{We can also consider more general cases with lattice site on a random graph or different Hilbert spaces on each lattice site. But in this paper we  start with the simplest case.} Thus the full system has a total Hilbert space that is a tensor product of the Hilbert space on each lattice point. 
We label the Hilbert space of the spin system at position $x$ to be $\ch_x$ and the state vector is labelled as $|i\>_x$ where $i=1,\ldots, d(\ch_x)\equiv d_x$. 
For simplicity, we consider all the Hilbert spaces to be identical. One simple example is a theory of $N$ pairs of free complex fermions $\psi^i, \bar{\psi}^i$, $i=1,\ldots, N$  on each site, so that the dimension of the Hilbert space is $d_x=2^N$. We can choose the ground state $|0\>$ to be annihilated by $\psi^i$ so that the states in the Hilbert space is spanned by $\bar{\psi}^{i_1}\ldots \bar{\psi}^{i_k}|0\>$. The simplest example is $N=1$ and the Hilbert space is just a single quantum bit whose dimension is $2$ and we can conveniently label them to be $|0\>$ and $|1\>$. We define a number operator at each site 
\bal
N_x = \bar{\psi}_x \psi_x \ .
\eal
It is clear that 
\bal
N_x|i\>=i |i\>\,, \qquad i = 0,1\ .\label{Ni}
\eal
We start with the system completely free, with neither on-site or inter-site interactions.

\subsection{Tracing over states}
Now we can turn on a source $\f_x$ conjugated to the number operator on each site $x$, the single site Hamiltonian then reads
\bal
H_x=m\bar{\psi}_x \psi_x-\bar{\psi}_x \psi_x  \f_x\ .\label{Hx}
\eal
Notice that at the moment we do not add any kinetic term to $\f_x$ and it is just a classical chemical potential. We will later consider the continuum limit where the kinetic term could emerge.
The theory describing this web of fermionic theories is defined by the Hamiltonian
\bal
H=\sum_x H_x\ .
\eal
Notice that in determining the dynamics of the system, we also need to provide the information about the quantum state of the spin.

Usually we consider the source to be classical and does not change much. But in the following, we will make this source a dynamical field. 
The first step to make it dynamical is to add conjugate momentum terms to the $\f_x$ on each site so that
\bal
H= \sum_x \left(\p_x^2+ H_x\right)\,,\label{ken1}
\eal
where $H_x$ is defined in~\eqref{Hx}.

We can in fact allow other terms involving only the $\f_x$ fields. Putting every thing together, we consider a  system described by the Hamiltonian
\bal
H=\sum_x H_{x,0} + H_{x,1}\,,\quad H_{x,0}=\p_x^2+ \frac{m}{2} \f_x^2+\hspace{-2mm}
\sum_y t_{xy}\f_x\f_y+ J_0(x)\f_x \,, \quad H_{x,1}=m\bar{\psi}_x \psi_x-\bar{\psi}_x \psi_x  \f_x\ .\label{Hall}
\eal

To proceed further, we allow the fermions to back react on $\f_x$ and try to find an effective theory of $\f_x$.
This can be done by tracing over the microscopic spin fields to get an effective action for $\f_x$. 
In the Hamiltonian formalism, 
the effective description of $\f_x$ is governed by the effective Hamiltonian $H_{\text{eff}}$ 
\bal
e^{-\b H_{\text{eff}}}=\text{STr}_{\ch}(e^{-\b H})=\text{Tr}_{\ch}((-1)^F e^{-\b H})=\text{Tr}_{\ch}( e^{-\b H+i \p F})\ .\label{Heffcorrect}
\eal
Notice that in addition to~\eqref{Heffcorrect}, there is another seemingly more natural definition of the $H_{\text{eff}}$
\bal
e^{-\b H'_{\text{eff}}}=\Tr_{\ch}(e^{-\b H})\,,\label{Heffwrong}
\eal  
which is the analogue of the partition function of the fermionic sector. But notice that this is not precisely the partition function since the theory also couples to the $\f$ field.
In addition, what we want is not the ``partition function" of the fermionic system, instead we want to integrate out the fermionic fields in a basis independent manner. For this reason~\eqref{Heffcorrect} is a better definition. 
To see this, consider a general basis of the Hilbert space $|i\>$, on which an operator takes the form $T^i_j$. Under a general change of basis $|i\>\to |i'\>=A_{i}^j|j\>$, which might include changes that mix bosonic with fermionic components, the operator becomes $A^j_l T^i_j \left(A^{-1}\right)^k_i \equiv ATA^{-1}$. Then the ``supertrace" $\text{STr}(T)=(-1)^{|i|}T^i_i$ changes to
\bal
& \text{STr}\left(AT A^{-1}\right)=(-1)^{|k|}  A^j_k T^i_j \left(A^{-1}\right)^k_i=(-1)^{|k|} (-1)^{|i|+|k|} \left(A^{-1}\right)^k_i A^j_k T^i_j   =(-1)^{|i|}T^i_i=\text{STr}(T)\,,
\eal  
where $(-1)^{|k|}$ is the action $(-1)^F$ on the outgoing state and $|k|$ labels the oddity under fermionic number operator. The extra sign factor comes from moving $A^{-1}$ through the $(-1)^{F}$ in the trace.  
It is clear that equation~\eqref{Heffwrong} does not satisfy this condition, which justifies the definition~\eqref{Heffcorrect}.
\footnote{
	An alternative understanding of this choice is that it imposes a periodic boundary condition on the fermionic $\psi^i$ fields. Another interpretation is that an imaginary chemical potential $e^{i\p F}$ is turned on.}

We prepare the system to be in a  state such that 
the fermions on the different sites do not entangle with each other, therefore the density matrix  is a tensor product of the density matrices at each site.
We prepare the system on each site to be in a mixed state with a density matrix
\bal
\r=\r_\f\otimes \r_\psi\,,\qquad \r_\psi= \bigotimes_x \r_x\,,\qquad \r_x= (1-p(x)) |0\>_x\<0|+ p(x) |1\>_x\<1| \ .\label{dens}
\eal 
Next we would like to consider a ``continuum limit" or the large-$n$ limit
\bal
n\to \infty\,,
\eal
where $n$ is the total number of lattice points per unit volume. Notice that here we have implicitly embedded the lattice into an ambient space $\cm = \mathbb{R}^d$ and the volume in the above sentence refers to the volume measured in the ambient space. This ambient space places no role in the following discussion except for setting up a scale.  

\begin{figure}[t]
	\centering
	\includegraphics[width=0.95\linewidth]{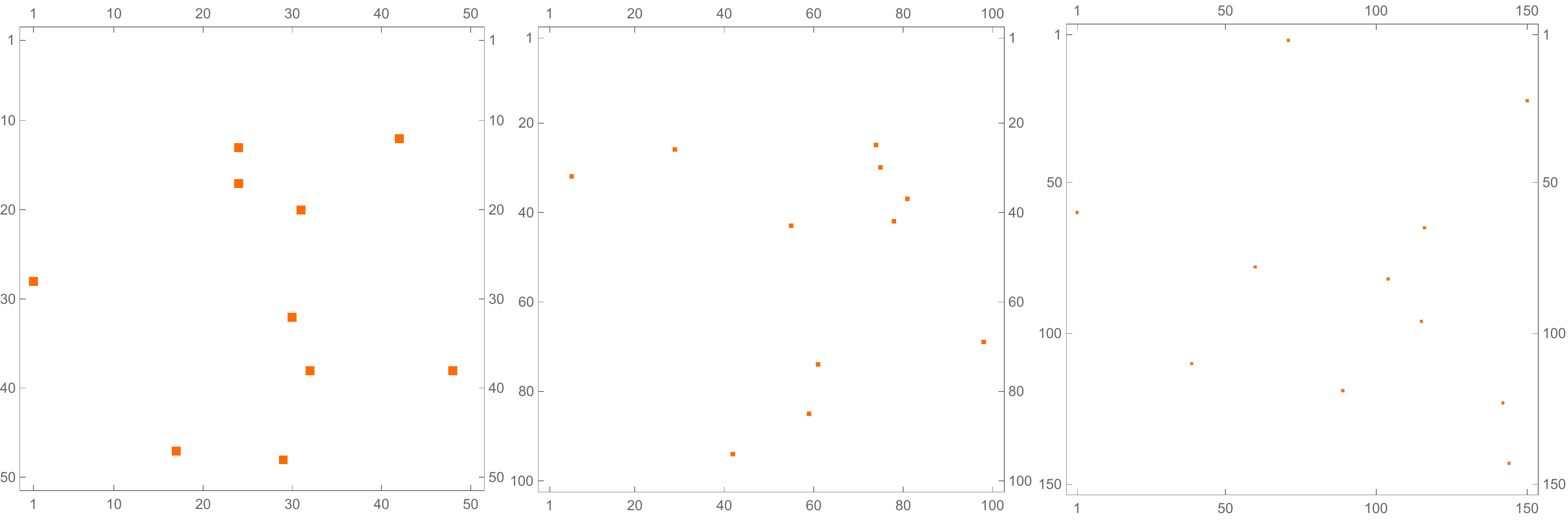}
	\caption{A numerical illustration of the limit~\eqref{limL}. The red dots are positions on the lattice where the state is in the $|1\>$ state. From left to right we have $n'=2.5\times 10^3 ,10^4, 2.25\times 10^4$ respectively while $\l(x)$ is fixed to 1 (with $dV(x)=10$ to make the figures easier to read). It is clear that as $n' \to \infty$ the expected number of excited states are approximately fixed.}
	\label{fig:pp1}
\end{figure}

We consider the limit where the total number of excitations are fixed and finite in this limit. This imposes a very non-trivial constraint on the probability function in the mixed state density matrix. 
\footnote{To satisfy this condition, there are two types of solutions. First, it could be that only at a finite number of sites the probability function $p(x)$ is finite, and the probability on all the others sites are zero. Second, it could also be that all the $p(x)$ on different site are of the same universal scale. Notice that the first solution is clearly not a RG fixed point, as we more and more coarse grain the sampling of the lattice the effective probability distribution will move towards the second solution. Therefore in the following we consider the second type of solution to the finite energy constraint. }
For simplicity we put the sites of the fermionic systems to be on a square lattice with the lattice parameter (i.e. intervals between each lattice point in each direction) $a$. 
Then the above continuous limit can be reached by taking $a=\frac{1}{m}$ with $m\to \infty$. It is clear that this grid of points is bijective to the grid of rational point in this limit. It is also clear that any open set on the background $\cm = \mathbb{R}^d$ contains a dense subset of points in the above $m\to \infty$ limit. Furthermore the density of grid point (per unit volume) is $n= m^d$. Therefore in the above continuous limit  there are $n V(B) \to \infty$ number of grid points in any open subset $B$, similar to the familiar fact that in any open subset of $\mathbb{R}$ there are countably infinite rational points. This fact remains true even if the volume of the open set is infinitesimal, such as the volume element $dV(x)$. 
The continuum limit is appropriately defined if the number of microscopic excitations in the fermion systems are finite 
\bal
\lim\limits_{n'\to \infty} n' p(x)=\L_{dx}(x)=\l(x)dV(x)\,, \qquad \l(x)\sim\co(1)\,,\label{limL}
\eal 
where we have indicated by $dx$ that the the $\L(dx)$ as a function of $x$ is also closely related to the set $dV(x)$. 
\footnote{As the name suggests, the $\L_{dx}(x)$ is related to the induced measure on the carrier space $\cm=\mathbb{R}^d$ in the previous Poisson process approach, and later we will comment that this newly induced ``dynamical" measure can be interpreted as describing an emergent gravity theory.}
We can thus factor out the $dV(x)$ dependence on the both sides to get
\footnote{Notice that in principle there is still the $dV(x)$ dependence in $\l(x)$ due to the different choice of the representative point in $dx$. But as we have assumed $p(x)$ to be smooth enough, this dependence drops out. In the following, we will always assume this and drop the $dx$ label in $\l(x)$. }
\bal
\lim\limits_{n\to \infty} n p(x)=\l_{dx}(x)\equiv \l(x)\,,\label{liml}
\eal
 An illustration of this limit is shown in figure~\ref{fig:pp1}.
 
Consider the fermionic systems on the sites inside the element $dV(x)$. We assume the $p(x)$ to be smooth enough so that in the small region $dV(x)$ it is approximately constant. 
   The fermionic factors in the density matrix that comes from the region $dV(x)$ is
   \bal
   \bigotimes_{x\in dV(x)} \r_x = \bigotimes_{x\in dV(x)} (1-p(x)) |0\>_x\<0|+ p(x) |1\>_x\<1|\ .
   \eal  
  In the particle number basis in $dV(x)$ it reads
   \bal
   \r_{dV(x)}&=\bigotimes_{x\in dV(x)} \left((1-p(x)) |0\>\<0|+ p(x) |1\>\<1|\right)\\
   &=\left((1-p(x)) |0\>\<0|+ p(x) |1\>\<1|\right)^{n'}\\
   &=P(n_x=k) \left(|0\>\<0|\right)^{\otimes (n'-k)}\otimes\left(|1\>\<1|\right)^{\otimes k} \,,\label{rhodv}
   \eal
   where as above $n'=n dV(x)$ is the total number of grid points in the volume element, $n_x$ is the total number of sites that are in the $|1\>$ states in the volume element $dV(x)$, equivalently it is the eigenvalue of the operator $\sum_{x\in dV(x)} N_x$. From this we read out the probability to have $n_x=k$ fermionic excitations
   in this volume is 
   \bal
    P(n_x =k) = {n'\choose k} p(x)^k (1-p(x))^{n'-k}\ .\label{pdv}
   \eal

In the limit~\eqref{limL}, this probability
becomes 
\bal
\lim\limits_{n'\to \infty} P_{dx}(n_x =k) &= \frac{n'!}{k!(n'-k)!n'^k} (n' p)^k (1-p)^{n'-k}\\
&=\frac{n'!}{k!(n'-k)!n'^k} (n' p)^k (1-\frac{n'p}{n'})^{n'-k}\\
&=\frac{1}{k!} (\L_{dx}(x))^k (1-\frac{\L_{dx}(x)}{n'})^{n'}\\
&=\frac{\L_{dx}(x)^k}{k!}  e^{-\L_{dx}(x)}=P_{\text{Pois}}(\L_{dx}(x),k)\ .\label{Pois1}
\eal
which means in the limit~\eqref{limL} the distribution of excitations in any open subset $dV(x)$ obeys a Poisson distribution.

Next we would like to trace over the Hilbert space of the spin degrees of freedom. For reasons discussed above, we consider the following trace
\bal
\text{Tr}_{\ch}\left(\r(-1)^F e^{-\b H}\right)=\text{Tr}_{\ch}\left(\r e^{-\b\left( H-\frac{i\p}{\b}F\right)}\right)=\text{Tr}_{\ch}\left(\bigotimes_{dV(x)}\r_{dV} \, e^{-\b\sum_x\left( H_{x,0}+H_{x,1}-\frac{i\p}{\b}F_x\right)}\right)\,,
\eal
Because $e^{i \p F}|n\>=e^{i \p N}|n\>$ and the bosonic fields $\f_x$ do not act on these fermionic component of the state (so they should be considered as functions in this computation), the $H_{x,1}$ and $F_x$ are diagonal on the states described by the density matrix. This together with~\eqref{rhodv},~\eqref{pdv} and~\eqref{Pois1} reduces the trace to 
\bal
&\text{Tr}_{\ch}\left(\bigotimes_{dV(x)}\r_{dV} \, e^{-\b\sum_x\left( H_x-\frac{i\p}{\b}F_x\right)}\right)=\text{Tr}_{\ch}\left(\bigotimes_{dV(x)}\r_{dV} \, e^{-\b\sum_x\left( H_{x,0}+(m-\f_x-\frac{i\p}{\b})N_x\right)}\right)\label{tr1}\\
&=\sum_{k_x=0}^{\infty} \left(\prod_{dV(x)} P_{\text{Pois}}(\L_{dx}(x),n_x = k_x)\right) e^{-\b \sum_{dV(x)}\left( H_{x,0}dV(x)+k_x (m-\f_x-\frac{i\p}{\b})\right)}\label{tr2}\\
&=e^{-\sum_{dV(x)} \left(\beta H_{x,0}dV(x)+\L_{dx}(x) e^{-\b \left( (m-\f)\right)} +\L_{dx}(x)\right)}\label{tr3}\\
&=e^{-\b {\bigintssss} \left(H_{x,0}dV(x)+\frac{\L_{dx}(x)}{\b}  e^{-\b \left( m-\f\right)} +\frac{\L_{dx}(x)}{\b}\right)}=e^{-\b {\bigintssss}\left(H_{x,0}+\frac{\l(x)}{\b} \left(  e^{-\b \left( m-\f\right)} +1\right)\right)dV(x)}\,,\label{tr4}
\eal
where in the last line we have taken the continuous limit and we have also used~\eqref{limL} and 
\bal
\sum_{n=0}^{\infty} \frac{\l^n}{n!} e^{-\l} e^{u n +c}= e^{\l (e^u-1)+c}\ .
\eal
In addition, the discretization is fine enough so that the $\f_x$ is approximately constant in each element $dV(x)$. 

 In addition, we notice that in the above derivation we did not specify any details about the decomposition of $\cm$ into the countable sum of open sets $dV(x)$. First, it is trivial to show the existence of such decompositions since the lattice points are countable and each open set contains a large number of lattice points, we can therefore use any lattice point in a given open set $dV(x)$ to label it. Since all the open set are disjoint, there is no ambiguity, such as repeated labelling, in this process. This shows the  existence of the decomposition. Second, the fact that the above derivation does not depend on the details of the decomposition simply means for different choices of the decomposition the result is always the same.

What  we have done so far is to consider a special scaling limit~\eqref{liml} of a lattice system, where a Poisson process description of the system is available. To see this clearly, we start with~\eqref{tr1} where we trace over the fermions. In the limit~\eqref{limL} this trace can be cast into the form of~\eqref{tr2} where we show that the number of excited lattice point in subset $dV(x)$ are mutually independent since  there is no fermionic hopping terms among different site. Furthermore the probability of $n_x$ in the mixed state represented by the density matrix~\eqref{dens} in the limit~\eqref{liml} is a Poisson distribution with Poisson parameter $\L(x)$. These together mean that the limit can be described by a Poisson process with the $n_x$ being the counting measure~\eqref{nb} and the $n' p(x)$ and $n p(x)$ quantities in the special limit~\eqref{liml} being the mean measure and the intensity of the Poisson process.
Therefore, as discussed in previous sections~\ref{randmath} and~\ref{randphy} this theory, in particular equation~\eqref{tr1}, can be considered as an ensemble of theories with a source $N_x$ that is related to a discrete Poisson distribution that is of the same type of the theory defined by~\eqref{freeaction} and~\eqref{aveerage}. But on the other hand, the theory clearly has another microscopic description as explained in this section. This then setup an equivalence with the ensemble averaged Poisson random theory discussed in section~\ref{randphy} and section~\ref{randmath} and gives an explicit example that an ensemble average of theories could actually be equivalently an average of an ensemble of states in a single (microscopic) theory.   
\footnote{Notice that the Poisson process discussed here is not to be confused with that discussed in section~\eqref{randmath} because here we work in the Hamiltonian formalism where the lattice is only in spatial directions.
 } 
 
 With this understanding, we actually do not have to do the above computation as in~\eqref{tr1}-\eqref{tr4}, rather this is nothing but the Laplacian functional of a Poisson process where the mean measure is identified with the $\L_{dx}(x)$ function in the continuum limit.    
Therefore we get an effective potential of the bosonic $\f$ source function as
\bal
\ch(x)=\frac{\l(x)}{\b}  e^{-\b \left( m-\f\right)} +\frac{\l(x)}{\b}=\frac{\l(x)}{\b} e^{-\b m}  e^{\b \f_x} +\frac{\l(x)}{\b}\ .
\eal
Defining
\bal
b=\frac{\b }{2}\,,\qquad  \m= \frac{\l}{2\p \b} e^{-\b  m}=\frac{\l}{4\p b} e^{-2b m}\,,\label{liouvpara}
\eal
we find the effective  potential to have the form of a (generalized) Liouville potential
\bal
\ch(x)=2\p \m(x) e^{2 b\f_x} +\frac{\l(x)}{2b}\ . \label{liouv}
\eal
Notice that explicitly the Hamiltonian is not exactly of the Liouville form since now the $\m$ is a function rather than a constant. This is similar with the results in the previous section~\ref{randphy} and~\ref{randmath}. Again we defer a more detailed discussion of the position dependent case in section~\ref{gravity}.

\subsection{The low energy limit}

  Next we focus on the low energy modes which are expected to have a canonical kinetic term in continuous spacetime and are exactly described by a 
  Liouville theory.
  This amounts to go to the frequency space and extract the effective action near zero momentum. 
  
In the following, we consider the simplest case with 1 spatial dimension. Then the position label $x$ in $\f_x$ is equivalently labelled by the order of the $\f_x$ on the chain, namely $x=j a$ where $j\in \mathbb{Z}$ and $a$ is just a scale of the grid interval. Equivalently, we can use this $j$ index to label the different fields. 

Our main consideration is on the following quadratic terms in~\eqref{Hall} 
\bal
H= \frac{m}{2}\sum_{j} \f_j^2+\sum_{|j-k|=1}t_{jk}\f_j\f_k\,,\label{H2}
\eal
where 
\bal
t_{ij}=t_{ji}\ .\label{sym}
\eal

For simplicity we consider a homogeneous chain so that the only non-vanishing hopping coupling is
\bal
t_{jj+1}=t\ .\label{simp1}
\eal

It is well known that a change of variables
\bal
\f_k \to \f'_k=\sqrt{\frac{1}{N+1}}\sum_{j}  \sin \left(\frac{\pi  (j+N+1) k}{2(N+1)}\right)\f_j \,,\qquad  k=1,\ldots, 2N+1\label{ffp}
\eal
diagonalizes the Hamiltonian~\eqref{H2}. 
Furthermore, it is instructive to define a spatial momentum
\bal
p_k= \frac{k \p}{2(N+1)a}\ .
\eal

In terms of this momentum the new field can be written as
\bal
\f'_k=\sqrt{\frac{1}{N+1}} \sum_{j}  \sin \left(p_k (x_j-x_0)\right)\f_j \,,\label{chg2}
\eal
where $x_0=-(N+1)a$.
From this expression it is evident that the $p_k$ has the meaning of a momentum since the form of the change of variable~\eqref{chg2} is in fact a discrete Fourier transform with the momentum space represented by $p_k$. 

With this new variable the diagonalized Hamiltonian then reads
\bal
H=\sum_{k}  \e_k{\f'}_k^2\,,\qquad \e_k=\frac{m}{2}-t \cos\left(\frac{k\p}{2(N+1)}\right)=\frac{m_\f}{2}-t \cos\left(p_k a\right)\ .
\eal
In the continuum limit $a \to 0, ~N a\to \infty$, so that $p_k \to 0$, the dispersion relation becomes
\bal
\e_k = \frac{m}{2}-t+\frac{t}{2}\left(p_k a\right)^2+\co(p_k^{4})\ .
\eal
In particular, if we choose
\bal
m=2m_\f^2+2t\,,\qquad  t=\frac{2}{c^2}a^{-2}\geq 0\,,\qquad m_\f\,, a\,, c \in \mathbb{R} \label{simp2}
\eal 
with $c$ being a constant, we recover the dispersion relation 
\bal
\e_k =m_\f^2+\frac{1}{c^2} p_k^2 + \co(p_k^4)\ .
\eal
Therefore, to the leading order we get
 with the simplification~\eqref{simp1} and~\eqref{simp2} we recover the spatial momentum term of a relativistic particle in a flat background. Explicitly, the Hamiltonian is
 \bal
 H_{sp}&=\frac{1}{c^2}\sum_k p_k^2 {\f'_k}^2 \\
 &=\frac{1}{c^2}\sum_k p_k^2 \left(\sqrt{\frac{1}{N+1}}\sum_{j}  \sin \left(p_k (x_j-x_0)\right)\f_j\right)^2\\
 &=-\frac{1}{c^2 (N+1)}\sum_{k=1}^{2N+1}  \left( \sum_{j=-N}^{N}  \sin \left(p_k (x_j-x_0)\right)\f_j\right)\pa_x^2\left( \sum_{j}  \sin \left(p_k (x_j-x_0)\right)\f_j\right)\\
 &=-\frac{1}{c^2 (N+1)}\pa_x^2\sum_k   \sum_{j=-N}^{N}  \sum_{r=-N}^{N}   \sin \left(p_k (x_j-x_0)\right) \sin \left(p_k (x_r-x_0)\right)\f_j \f_r\\
  &=-\frac{1}{c^2 (N+1)}\pa_x^2   \sum_{j=-N}^{N}  \sum_{r=-N}^{N}  (N+1) \delta_{j,r}\f_j \f_r\\
  &=-\frac{1}{c^2 }   \sum_{j=-N}^{N}   \f_x \pa_x^2\f_x =\frac{1}{c^2 }   \sum_{j=-N}^{N}    \left(\pa_x\f_x\right) ^2\label{ken2}
 \eal
where we have used~\eqref{chg2}. In the third line we have regarded the $x_j$ formally as the discrete value of a continuous position coordinate with respect to which the $\pa_x$ operator is defined. It should be considered as the continuous limit of the difference operator. 

Therefore together with~\eqref{Hx},~\eqref{ken1},~\eqref{liouvpara},~\eqref{liouv},~\eqref{H2} and~\eqref{ken2}, we get the following action for the low energy continuous limit of the theory obtained from tracing over the microscopic fermionic degrees of freedom
\bal
H=\sum_x \left(\p_x^2+\frac{1}{c^2 }    \left(\pa_x\f_x\right) ^2+ J_0(x) \f_x+2\p \m(x) e^{2 b\f_x} +\frac{\l(x)}{2b}\right)\,,\label{lioumic}
\eal
where we have taken the $n\to \infty$ limit of~\eqref{ken2}. We find this is the Hamiltonian of the Liouville theory, similarly as that from averaging over a Poisson distributed random sources.

Further notice that in this derivation, we have tuned the parameters in the microscopic model so that in the continuum limit the scalar fields has a classical relativistic kinetic term on flat Euclidean spacetime. We could in fact consider more general parameters in the microscopic model, which will lead to a more general kinetic term.

\section{Sinh-Gordon type models}

In the previous example, the random source is drawn from a Poisson process which has the property that only positive values are supported in the distribution. One would ask if this discussion is only specific to the Poisson distribution or can be made more general. In addition, one would naturally ask what if we consider a distribution where the random variable can take both positive and negative (discrete) values, which is more similar to the Gaussian distribution. 
In the following we consider such a model from the random ensemble average point of view and the microscopic points of view. We will show that the average of this model leads to Sinh-Gordon type theories.

\subsection{The random ensemble point of view}

First, on the level of random average of different theories, we could consider a different Probability distribution of the source. 
For example, we can consider a similar model of
\bal
\cl = \pa_\m \f \pa^\mu \f - \left(J_0(x) + J_1(x)\right) \f\,,
\eal
where $J_1(x)$ is related to the Skellam distribution~\eqref{sk1}
\bal
P(x;\mu_1,\mu_2)= e^{-(\mu_1+\mu_2)}
\left(\frac{\mu_1}{\mu_2}\right)^{x/2}I_{x}(2\sqrt{\mu_1\mu_2})\,,\qquad~~ x\in \mathbb{Z}\,,
\eal
where $I_x$ is the modified Bessel function of the first kind and the $\m_1$ and $\m_2$ are parameters characterizing this random distribution
\bal
\<J_1\> = \m_1-\m_2\,, \qquad  \<J_1^2\>-\<J_1\>^2 = \m_1 +\m_2\,, \qquad \m_1,\m_2 \geq 0 \ .
\eal

The random averaging can be done following the above Liouville discussion in either the more physical way discussed in section~\ref{randphy} or the more mathematical way in section~\ref{randmath}. In the latter approach, the random sources can still be considered from a point process, the only difference is that now the probability distribution of the $N(B)$ for any open set on $\cm$ is a Skellam distribution. Then following a similar discussions, we arrive at the solution with the following effective interaction Hamiltonian   
\bal
 \ch(x)= \frac{1}{\b}\left(\mu_1(x)+\mu_2(x)+\mu_1(x)e^\F+\mu_2(x)e^{-\F}\right)\label{Hsk}
\eal
where $\m_1(x)$ and $\m_2(x)$ are two functions parametrizing the point process with the Skellam distribution measure on different open sets.

\subsection{The microscopic point of view}

As the above Poisson process case, there is again a microscopic setting which realizes the above theory as an average over an ensemble of microstates in a single theory. This provides another example showing that averaging over different theories could be understood as averaging over an ensemble of states within one  theory in an appropriate limit. 

Explicitly, we consider a microscopic model with again a grid of sites on which some fermionic quantum mechanical modes exist. We consider a simplest model with two complex fermions $\psi^1_x$ and $\psi^2_x$ on each site labelled by $x$. The interaction Hamiltonian of the collection of the fermionic systems reads
\bal
H_x(t) &= \bar{\psi}^1_x(t) \psi^1_x(t)\f_x(t)-\bar{\psi}^2_x(t) \psi^2_x(t)  \f_x(t)\ .
\eal 
Notice that there is a relative minus sign between the two terms, this reflects the fact that we assign opposite counting charges of the two fermions under the particle number operator, mimicking the electron-hole pair in more familiar systems. 

One can again consider the microstates to be in an ensemble  represented by a factorized density matrix
\bal
\r_\psi= \r_{\psi^1}\otimes\r_{\psi^2}\,,
\eal
where
\bal
\r_{\psi^i}= \bigotimes_{x,i} \r_{x,i}\,, \qquad 
\r_{x,i}= (1-p_i(x)) |0\>_{x,i}\<0|+ p_i(x) |1\>_{x,i}\<1|\ .
\eal 
Here we emphasize again that this is only a special mixed state that we choose the microscopic fermions to stay in. The system could very well be in a different mixed states, and in those cases the effective theory of $\f$ could be very different. 

The Hamiltonian is closed related to the ``net" number operator 
\bal
M_x=\bar{\psi}^1_x  \psi^1_x -\bar{\psi}^2_x {\psi}^2_x  \,,
\eal
which satisfies
\bal
M_x|0\>_{x,i}=0\,,\qquad  M_x|1\>_{x,i}=(3-2 i)|1\>_{x,i}\ .
\eal

Next we would like to consider the ``continuum limit" in the same sense as the in previous Liouville discussion in section~\ref{micro}
\bal
n\to \infty\,,
\eal
where $n$ is the total number of lattice points per unit volume. 
Concretely,  we again assume $p_i(x)$ to be smooth enough so that in the small region $dV(x)$ it is approximately a constant, so we consider the limit 
\bal
\lim\limits_{n'\to \infty} n' p_i(x)=\L_{i,dV}(x)\,,\label{limL2}
\eal 
and similarly
\bal
\L_{i,dV}(x)=\m_i(x)dV(x)\,, \qquad \m_i(x)\sim\co(1)\ .\label{measureL2}
\eal
Once again, we could consider $\L_{i,dV}(x)$ as an induced measure on the carrier space $\cm=\mathbb{R}^d$. 

Then the probability of the total net excitation in this volume element $dV(x)$ being $m_{dV}=k$ with $k>0$ is 
\bal
&P(m_{dV}=k) =\sum_{n_2=\max(0,-k)}^{\infty} P(n_1 -n_2 =k) \\
&= \sum_{n_2=\max(0,-k)}^{\infty}{n'\choose n_2+k} p_1(x)^{n_2+k} (1-p_1(x))^{n'-n_2-k}{n'\choose n_2} p_2(x)^{n_2} (1-p_2(x))^{n'-n_2}\,,
\eal 
which can be simply counted using the fact that the systems on each sites are independents. We have also only computed the $k>0$ case, for $k<0$, we can rewrite the sum in terms of $n_1=n_2+k$ and the result is the same as $P(m_{dV}=-k)$. 

In the limit~\eqref{limL2}, the above probability becomes 
\bal
&\lim\limits_{n'\to \infty} P(m_{dV}=k) \\
&=  \sum_{n_2=\max(0,-k)}^{\infty}\frac{\L_{1,dV}(x)^{n_2+k}}{(n_2+k)!}  e^{-\L_{1,dV}(x)}\frac{\L_{2,dV}(x)^{n_2}}{(n_2)!}  e^{-\L_{2,dV}(x)}\\
&=  \sum_{n_2=\max(0,-k)}^{\infty}P_{\text{Pois}}(\L_{1,dV}(x),n_2+k)P_{\text{Pois}}(\L_{2,dV}(x),n_2)\\
&= e^{-(\L_{1,dV}(x)+\L_{2,dV}(x))}
\left(\frac{\L_{1,dV}(x)}{\L_{2,dV}(x)}\right)^{k/2}I_{k}(2\sqrt{\L_{1,dV}(x)\L_{2,dV}(x)})\,,
\eal
where $I_k(x)$ is the  modified Bessel function of the first kind. For $k<0$ we follow the same procedure, and the result can be put into a uniform expression for all values of $k$
\bal
&\lim\limits_{n'\to \infty} P(m_{dV}=k):=P_{\text{Sk}}(\L_{1,dV}(x),\L_{2,dV}(x),k)\\
&\quad = e^{-(\L_{1,dV}(x)+\L_{2,dV}(x))}
\left(\frac{\L_{1,dV}(x)}{\L_{2,dV}(x)}\right)^{k/2}I_{k}(2\sqrt{\L_{1,dV}(x)\L_{2,dV}(x)})\ .\label{sk2}
\eal
This probability on the open set $dV(x)$ is nothing but the Skellam distribution with the mean value $\L_{1,dV}(x)-\L_{1,dV}(x)$ and the variance $\L_{1,dV}(x)+\L_{1,dV}(x)$, both depending on the position $x$ and the open set $dV(x)$.

Next we would like to trace over the Hilbert space of the Fermi system, which is equivalent to integrating out the spin system background. We again consider the following trace
\bal
\text{Tr}_{\ch}\left(\r(-1)^F e^{-\b H}\right)=\text{Tr}_{\ch}\left(\r e^{-\b\left( H-\frac{i\p}{\b}F\right)}\right)=\text{Tr}_{\ch}\left(\bigotimes_{dV(x)}\r_x e^{-\b\sum_{dV}\left( H_{dV(x)}-\frac{i\p}{\b}F_{dV(x)}\right)}\right)\,,
\eal
which follows from a similar derivation as in the previous Liouville theory case. For simplicity we drop the terms that only depend on the bosonic field. As shown in the previous case, those terms do not affect the evaluation of the trace and can be put back at the end.
The evaluation of this trace reads
\bal
&\text{Tr}_{\ch}\left(\bigotimes_{dV(x)}\r_x e^{-\b\sum_{dV}\left( H_{dV(x)}-\frac{i\p}{\b}F_{dV(x)}\right)}\right)\\
&=\sum_{k_x}\left(\prod_{dV(x)} P_{\text{Sk}}(\L_{1,dV}(x),\L_{2,dV}(x),k) e^{-\b \sum_{dV(x)}\left( k (m+\f_x-\frac{i\p}{\b})\right)}\right)\\
&=e^{\sum_{dV(x)}\left(\L_{1,dV}(x) \left(-e^{-\b \left( (m+\f)\right)}-1\right) +\L_{2,dV}(x) \left(-e^{\b \left( (m+\f)\right)}-1\right)\right)}\\
&=e^{-\b {\bigintssss}dx  \left( \frac{\m_1(x)}{\b} e^{-\b \left( m+\f\right)}+\frac{\m_2(x)}{\b} e^{\b \left( m+\f\right)} +\frac{\m_1(x)}{\b}+\frac{\m_2(x)}{\b}\right)}\,,
\eal
where we have used $e^{i \p F}|n_1,n_2\>=e^{i \p (n_1+n_2)}|n_1,n_2\>=e^{i \p (|n_1-n_2|)}|n_1,n_2\>$ and also we have adopted the identification of $\L_{i,dx}(x)=\m_i(x)dx$ as an induced measure to rewrite the weighted sum in the exponential into an integral over $\cm$ with a dynamical measure. 

We thus find the resulting Hamiltonian is of the form of a Sinh-Gordon type potential. We can follow the same treatment in the previous section~\ref{micro} to show the emergence of a relativistic kinetic terms and obtain the full  Hamiltonian of the Sinh-Gordon action.

\section{Discussion}

In this section we discuss some interesting questions and extensions related to the material in the above main text.
  
  \subsection{Relation to Gaussian randomness}
  
  Focusing on discrete distributions is a crucial difference between this work and the previous literature on ensemble averages of continuous Gaussian type random variables.   It is often the case that a further limit of the discrete distribution, for example the Poisson distribution that we mainly discussed here, gives a Gaussian distribution, which is guaranteed by the central limit theorem as long as the events that the distribution describes are mutually independent. Therefore we can also regard the discussion in this paper as a first attempt towards refining the results in previous Gaussian random (holographic) theories, including SYK type models and the random matrix theories, in the sense of reverting the limit to go back from Gaussian to discrete distributions.
  
\subsection{The quantum mechanics dual description}
We can also integrate out the $\f_x$ fields and ask what does the resulting microscopic model look like. We expect the result, which is a 0+1d quantum mechanical model with $x$ being discrete flavour labels of the fermions, to be a purely field theoretical description of the same system. 

Explicitly, it is easier to work in the Lagrangian formalism, where the coupled theory is defined as
\bal
L^E=\sum_x\left(\frac{1}{2}\dot{\f_x}^2+\frac{m_\f}{2}\f_x^2+\frac{t_{x-1,x}}{2}\f_{x-1}\f_x+\frac{t_{x,x+1}}{2}\f_x\f_{x+1}+i\bar{\psi}\dot{\psi}+m  \bar{\psi}_x\psi_x-\bar{\psi}_x\psi_x\f_x\right)\ .
\eal 
Notice that in this section we again work in Euclidean signature.
Treating $x$ as a flavour index, we can integrate out the $\f_x$ to get
\bal
e^{- \int dt L^E_{eff}}&= \int \cd \f_x e^{-\int dt L^E}\sim e^{-\frac{1}{2}\log(M)+\int dt \left(i\bar{\psi}\dot{\psi}+m  \bar{\psi}_x\psi_x+\frac{1}{2}\bar{\psi}_x\psi_x\bar{\psi}_y\psi_y(M^{-1})_{xy}\right)}\,,
\eal
where $\sim$ simply means up to an irrelevant constant and the matrix $M$ is defined to be
\bal
M_{xy}=-\pa^2_\t\delta_{x,y}+m_\f\delta_{x,y}+t_{x-1,x}\delta_{y,x-1}+ t_{x,x+1}\delta_{y,x+1}\ .\label{Mxy}
\eal
We observe that the quantum mechanical description of the same model is characterized by a complex fermion system with a ``nearest neigh bore" charge-charge coupling. 

We can further consider the $M^{-1}$ factor as the coupling constants of the charge-charge interaction. This interaction can be analysed by a derivative expansion of the inverse of~\eqref{Mxy}. In particular, the leading term in this expansion is simply the inverse of~\eqref{Mxy} with the derivative term turned off. The next-to-leading term is a 2-derivative term, which is irrelevant in the quantum mechanics system. As a result, for the purpose of obtaining an low energy effective theory we can simply drop all the derivative corrections to get the following effective action
\bal
S_{\text{eff}}=-\int dt \left(i\bar{\psi}\dot{\psi}+m  \bar{\psi}_x\psi_x+\frac{1}{2}g_{xy}\bar{\psi}_x\psi_x\bar{\psi}_y\psi_y \right)\,,
\eal 
where
\bal
g_{xy}&=\left(m_\f\delta_{x,y}+t_{x-1,x}\delta_{y,x-1}+ t_{x,x+1}\delta_{y,x+1}\right)^{-1}\ .
\eal
We can again consider the special value~\eqref{sym} and~\eqref{simp1} where the inverse can be computed as
\bal
g_{xy}&=\left(m_\f\delta_{x,y}+t\delta_{y,x-1}+ t\delta_{y,x+1}\right)^{-1}\\
&=\frac{1}{m_\f}\sum_{k=0}\left(\frac{-t}{m_\f} (\delta_{y,x-1}+\delta_{y,x+1})\right)^k\\
&=\frac{1}{m_\f}\sum_{k=0}\left(\frac{-t}{m_\f}\right)^k \sum_{j=0}^{k}{k\choose j}\delta_{y,x+k-2j}\ .
\eal
Clearly $g_{xy}=g_{yx}$ so we can focus on $y\geq x$ cases where the sum over $j$ reduces to half of the range
\bal
g_{xy}&=\frac{1}{m}\sum_{k=0}\left(\frac{-t}{m}\right)^k \sum_{j=0}^{\lfloor \frac{k}{2}\rfloor }{k\choose j}\delta_{y,x+k-2j}\\
&=\frac{1}{m}\sum_{k=0}\left(\frac{-t}{m_\f}\right)^{2k} \sum_{j=0}^{k}{2k\choose j}\delta_{y,x+2k-2j}+\frac{1}{m}\sum_{k=0}\left(\frac{-t}{m}\right)^{2k+1} \sum_{j=0}^{k }{2k+1\choose j}\delta_{y,x+2k+1-2j}\\
&=\frac{1}{m}\sum_{p=0}^{\infty }\sum_{k=p}^{\infty}\left(\frac{-t}{m}\right)^{2k} {2k\choose k-p}\delta_{y,x+2p}+\frac{1}{m}\sum_{p=0}^{\infty }\sum_{k=p}^{\infty}\left(\frac{-t}{m}\right)^{2k+1} {2k+1\choose k-p}\delta_{y,x+2p+1}\\
&=\frac{1}{m}\sum_{p=0}^{\infty }\sum_{k=0}^{\infty}\left(\frac{-t}{m}\right)^{2k+2p} {2k+2p\choose k}\delta_{y,x+2p}+\frac{1}{m}\sum_{p=0}^{\infty }\sum_{k=p}^{\infty}\left(\frac{-t}{m}\right)^{2k+2p+1} {2k+2p+1\choose k}\delta_{y,x+2p+1}\\
&=\frac{1}{m \sqrt{1-\frac{4 t^2}{m^2}}}\sum_{p=0}^{\infty }\left(\frac{ \left(\frac{2t}{m}\right)^{2 p} }{\left(\sqrt{1-\frac{4 t^2}{m^2}}+1\right)^{2 p}}\delta_{y,x+2p} -\frac{ \left(\frac{2t}{m}\right)^{2 p+1} }{\left(\sqrt{1-\frac{4 t^2}{m^2}}+1\right)^{2 p+1}}\delta_{y,x+2p+1}\right)\ .
\eal
Adding back the $y<x$ terms, we simply get
\bal
g_{xy}=\frac{1}{m \sqrt{1-\frac{4 t^2}{m^2}}}\sum_{p=-\infty}^{\infty }\left(\frac{ \left(\frac{2t}{m}\right)^{2 |p|} }{\left(\sqrt{1-\frac{4 t^2}{m^2}}+1\right)^{2 |p|}}\delta_{y,x+2p} -\frac{ \left(\frac{2t}{m}\right)^{2 |p|+1} }{\left(\sqrt{1-\frac{4 t^2}{m^2}}+1\right)^{2 |p|+1}}\delta_{y,x+2p+1}\right)\ .
\eal
For the special value~\eqref{simp2}, the parameter in the above result behaves as
\bal
\frac{t}{m}=\frac{1}{2(1+\frac{m_\f^2}{t})}\leq \frac{1}{2}\ .
\eal 
In this range, the coupling  considered above is in general real and positive. However, when the inequality is saturated,  the coupling actually diverges. 

The physical reason of this divergence is clear. When the inequality saturates $m_\f =0$ according to~\eqref{simp2}, so the effective mass of the infrared $\f$ mode vanishes. When this happens, integrating out this massless modes is inconsistent and  leads to  divergences. What we have observed is just a realization of this well known phenomenon, see e.g.~\cite{Seiberg:1994bz,Seiberg:1994rs} in our very simple model, where no gauge symmetry or supersymmetry is involved.   

Further notice that the results in this section are simplified version due to the assumption~\eqref{sym} and~\eqref{simp1}. Without these assumptions we expect to get richer structure of this dual description.

\subsection{Interpretation as emergent gravity ?}\label{gravity}

From the results of the previous sections, we observe that Liouville theory and generalizations could arise from averaging over an ensemble of theories with random sources. The latter type of theories could have a microscopic origin where one traces over a set of underlying fermionic degrees of freedom in a specific state to generate an effective theory of the rest degrees of freedom. 
In both the two scenarios a key point is that by relating such averaging or tracing out the microscopic degrees of freedom to the mathematical description of (Poisson) point processes, an emergent probabilistic measure on the carrier space naturally appears. 
One is then tempting to identify such an emergent measure, which ultimately comes from tracing over the fermionic microstates, with a geometric metric which facilitates an emergent gravity theory whose action takes the form of Liouville action. 

This does not sound that unreasonable given that the 2D Liouville theory itself does emerge from the response of the conformal matter fields to a Weyl transformation of the background metric, see e.g.~\cite{DHoker:1990prw}.  
The only difference in our setting is that we do not start with a conformal matter fields that couples to a gravitational background. Instead we start from a large number of microscopic fermionic modes that couples to a bosonic mode, then in a  double-scaled continuous limit an effective Liouville action emerges for the bosonic modes.

Concretely, recall that the effective action of the Liouville theory can be obtained from the response of the matter field to a Weyl transformation of the metric $h_{\m\n} \to e^{\f} h_{\m\n}$, with the effective action
\bal
I_{\mathrm{L}}^{(b)}=\frac{1}{4 \pi} \int \mathrm{d}^{2} x \sqrt{|h|}\left(Q \Phi(x) R_h(x)+(\nabla \Phi)^{2}+4 \pi \mu e^{2 b \Phi(x)}\right)\,,\label{liouv2}
\eal
where $\mu$ is a cosmological constant and $Q$ is a background charge related to the property, such as the central charge, of the matter fields.

In our analysis, the effective action of the free scalar theory with a Poisson random source has been shown~\eqref{Seff1} to be
\bal
S_{\text{eff}}= \int d^d x \left(\pa_\m \f \pa^\mu \f(x) - J_0(x) \f +\l(x)(e^{\f(x)}-1)\right)\ .
\eal
Given the resemblance of the two actions, we would like to try matching the two actions by identifying the dilaton $\F$ in~\eqref{liouv2} with the boson $\f$ in~\eqref{Seff1} for $d=2$. However, given the form of the two actions, it is the simplest if we fix a gauge of the gravitational theory with dynamical metric field $h_{\m\n}$. This can be directly observed from a simple comparison: fluctuations of the metric, not only its determinant, is dual to a change of the kinetic term that can be traced to a different hopping structure on the microscopic lattice. Thus choosing a different hopping structure on the lattice correspond to a change of the metric. Fixing the metric to a certain gauge is them mapped to a choice of the hopping terms on the lattice. For example, for the choice of the hoppings we discussed above, the effective kinetic term is simply the Laplacian on the conformally flat spacetime. To make the comparison transparent, we fix the metric in~\eqref{liouv2}  to the conformal gauge
\bal
h_{\m\n}=e^{\r(x)}\delta_{\m\n}\ .\label{gauge}
\eal

In this gauge, the above duality simplifies to
\bal
&-Q \delta^{\m\n}\pa_\m\pa_\n \r(x) = - J_0(x)\\
&4 \p \m  e^{\r(x)}=2 \p \m(x)=\frac{\l(x)}{2b}e^{-2mb } \\
&  \delta^{\m\n}\pa_\m\pa_\n =\delta^{\m\n}\pa_\m\pa_\n  \ .
\eal
The third line reduces to an identity as a result of our fixing the gauge~\eqref{gauge}, while the first two equations implies a condition between the canonical source $J_0(x)$ and the $\l(x)$ that can either be understood as the ``parameter" characterizing the point process or characterizing the state of in which the effective action describes in the microscopic ensemble average description. In particular, the relation is
\bal
\frac{J_0(x)}{Q}= \delta^{\m\n}\pa_\m\pa_\n \log\left(\frac{\l(x)e^{-2mb}}{8\p\m b}\right)=\delta^{\m\n}\pa_\m\pa_\n \log\left(\l(x)\right)\,,\label{cond}
\eal 
where in the last expression we have assumed that all the other parameters in the microscopic setting are independent of $x$.

This is a very interesting relation stating that we can consider the $J_0$ as also the source of the $\log(\l(x))$ when the latter is regarded as a classical field. This means for a given $J_0(x)$ the value of the function $\l(x)$ is determined once a boundary condition is provided. Microscopically, this relation has a clearer interpretation: for any given $J_0(x)$ the microscopic state in which the system has a gravitational description can be identified so that the density matrix of the state~\eqref{dens} scales correctly~\eqref{limL} so that the limiting $\l(x)$ function should satisfies the condition~\eqref{cond}.
This agrees with the general philosophy of gauge/gravity duality where the content of the duality depends crucially on which states the quantum field theory is in. It is true that the limit~\eqref{limL} and the condition~\eqref{cond} can not isolate a single state in the microscopic theory. But this is also as what we would expect since in our picture the different states  giving the same limit~\eqref{limL} is not distinguishable from the semi-classical gravity interpretation. It is likely that the different states with the same~\eqref{limL} carries information about quantum corrections to the classical gravity/geometry. We plan to understand this proposal better in the future.   

Further notice that the requirement of a gravitational interpretation only fix the shape dependence of $J_0(x)$ on $\l(x)$, the size of the $J_0(x)$ field can however vary and it is proportional to $Q$ in the gravity interpretation. We know that in the derivation of the Liouville theory as a response of matter fields to a Weyl transformation of the gravity background they couple to, $Q$ roughly counts the number of degrees of freedom of the matter fields, and hence an indication of the strength of  the matter coupling. In our derivation, the $\f(x)$ scalar field can be interpreted as the dilaton in the 2d gravity of the Liouville type, and then the size of $J_0(x)$ can thus be understood as the strength of the coupling to some other matter field that could be considered as other conformal matters fields in the gravity interpretation.

Of course this identification and interpretation are specialized in the conformal gauge~\eqref{gauge}. It is definitely an interesting question to set up a concrete relation covariantly. One simple idea is that one should be able to impose conditions on the allowed probability measure $\L(dx)=\l(x)dV(x)$ so that there is an emergent diffeomorphism on the $\l(x)$ that can be identified with the diffeomorphism in the gravity description  But this is beyond the scope of this paper and we will discuss this in more details in future work. Nevertheless, in either respect, namely either our result being  gauge dependence or in the near future we consider diffeomorphism invariant $\l(x)$ measure as the effective background metric, our construction and in particular the interpretation of emergent gravity is not forbidden by the Weinberg-Witten theorem.

Finally, we notice that  connection between dilaton gravity models and the Liouville-type theory have been discussed recently~\cite{Maldacena:2016hyu,Bagrets:2016cdf,Bagrets:2017pwq,Mertens:2017mtv,Mertens:2020hbs,Maxfield:2020ale}  and similar discussion about the relation between the dilaton gravity theory with the sinh-Gordon and other types of theory~\cite{StanfordSeiberg,Mertens:2020hbs,Witten:2020wvy}.  
There have also been some discussions about the connection between 1d Liouville and the Schwarzian theory. In the paper~\cite{Bagrets:2016cdf,Bagrets:2017pwq}, the author considered a rewriting of the Schwarzian theory as a Liouville theory with some regularization. In~\cite{Mertens:2017mtv} an indirect connection between the two theories are made via the reduction from integrating in and out different Lagrangian multiplier fields.

\subsection{Concluding remarks}

In this paper, we consider averaging over theories with discrete probability distribution, and show an explicit equivalence between this type of theories with the effective theory from tracing over part of the microscopic degrees of freedom of a single theory in some appropriate limit. 
Further notice that in this discussion, especially from the point of view of the Poisson point process, the spatial dimension of the carrier space is not crucial. Therefore it is possible to compare the above approach to some other known random deformations of free field theories in higher dimensions ~\cite{Berkooz:2016cvq,Murugan:2017eto,Berkooz:2017efq,Bulycheva2017,Peng:2018zap,Ahn:2018sgn,cepm}. 

While finishing the paper, we noticed other discussions of probabilistic construction of Liouville conformal field theory~\cite{2020arXiv200511530G}. The approach there is very different from ours; there the Liouville field itself is considered as a random field under a specific probability measure. In our analysis in section~\ref{randphy} and~\ref{randmath}, we only consider the source to be related to Poisson process, while the field $\f(x)$ itself is conventional. So at the moment we have not observed any direct connection between our analysis and the approach used in~\cite{2020arXiv200511530G} and the reference therein. It will be interesting to better understand possible connections in the future.

\vspace{3mm}
\centerline{\bf Acknowledgments}
\vspace{1mm}
We thank Eduardo Casali, Chi-Ming Chang, Liangyu Chen, Sean Colin-Ellerin, Temple He, Veronika Hubeny, Yang Lei, Yan Liu,  Mukund Rangamani, Ya-Wen Sun, Jia Tian, Huajia Wang, Jieqiang Wu, Yingyu Yang, Zekai Yu, Zhixian Zhu and other members at KITS for many interesting discussions on related topics. The work is supported by funds from the Kavli Institute for Theoretical Science (KITS) and a startup funding from the University of Chinese Academy of Science (UCAS), this work is also supported in part by the U.S. Department of Energy grant DE-SC0019480 under the HEP-QIS QuantISED program and by funds from the University of California.

\appendix

\section{Discrete distributions}

For the convenience of later computation it is useful to first list some properties of the distributions discussed in the main text.

$\bullet $ The Poisson distribution has the probability mass function
\bal
P(x,\l)=e^{-\l}\frac{\l^{x}}{x!} \Theta(x)\,, \qquad  x\in \mathbb{Z}\,, \label{ps1}
\eal
where 
\bal
\Theta(x)=\begin{cases}
	1\,, \qquad x\geq 0\\
	0\,, \qquad x<0
\end{cases} \ .
\eal

Its moment generating function (MGF) and characteristic function (CF) are
\bal
\text{MGF}:& M_x(\F)=\mathbb{E}[e^{\F x}]=\sum_{-\infty}^{\infty}  P(x,\l) e^{\F x} = e^{\l e^\F -\l}\label{mgps}\\
\text{CF}:& C_x(\F)=\mathbb{E}[e^{i\F x}]=\sum_{-\infty}^{\infty}  P(x,\l) e^{i\F x} = e^{\l e^{i \F} -\l}\,, \label{cfps}
\eal
where $\mathbb{E}$ is the mean value of the distribution assuming the random variable to be $x$.

$\bullet$ The Skellam distribution~\cite{skellam} has the probability mass function
\bal
P(x;\mu_1,\mu_2)= e^{-(\mu_1+\mu_2)}
\left(\frac{\mu_1}{\mu_2}\right)^{x/2}I_{x}(2\sqrt{\mu_1\mu_2})\,,\qquad  x\in \mathbb{Z}\,,\label{sk1}
\eal
where $I_k(x)$ is the  modified Bessel function of the first kind.
The MGF and CF are
\bal
\text{MGF}:& M_x(\F)=\mathbb{E}[e^{\F x}]=\sum_{-\infty}^{\infty}  P(x,\m_1,\m_2) e^{\F x} = e^{-(\mu_1+\mu_2)+\mu_1e^\F+\mu_2e^{-\F}}\label{mgsk}\\
\text{CF}:& C_x(\F)=\mathbb{E}[e^{i\F x}]=\sum_{-\infty}^{\infty}  P(x,\m_1,\m_2) e^{i\F x} = e^{-(\mu_1+\mu_2)+\mu_1e^{i\F}+\mu_2e^{-i\F}}\ .\label{cfsk}
\eal
To understand better the meaning of the Skellam distribution we consider two independent Poisson distributions $P(x_1,\m_1)$ and $P(x_2,\m_2)$. Then the joint distribution is
\bal
P(x_1,\m_1)P(x_2,\m_2)=e^{-\m_1}\frac{\m_1^{x_1}}{m_1!}e^{-\m_2}\frac{\m_2^{x_2}}{x_2!}\ .\label{prod}
\eal 
It is easier to consider directly the generating function
\bal
G(z;\l)=\sum_{x=-\infty}^{\infty} P(x,\l) z^x\ .
\eal
For the current Poisson distribution, this gives
\bal
G(z;\l)=\sum_{x=0}^{\infty} e^{-\l}\frac{\l^{x}}{x!} z^x=e^{\l z -\l}\ . 
\eal
Then the above joint distribution~\eqref{prod} is simply the $z_1^{x_1} z_2^{x_2}$ coefficient of
\bal
G(z_1,z_2;\l_1,\l_2)=G(z_1;\l_1)G(z_2;\l_2)=e^{\l_1 (z_1 -1)+\l_2(z_2-1)}=e^{-(\l_1+\l_2)+ (\l_1 z_1 +\l_2 z_2)}\ .
\eal
Consider the special case 
\bal
z_2= \frac{1}{z_1}\,,
\eal
where the generating function becomes
\bal
G(z_1,\frac{1}{z_1};\l_1,\l_2)=e^{-(\l_1+\l_2)+ (\l_1 z_1 +\l_2 z_1^{-1})}=e^{-(\l_1+\l_2)+\sqrt{\l_1\l_2}\left(\sqrt{\frac{\l_1}{\l_2}} z_1 +{\left(\sqrt{\frac{\l_1}{\l_2}} z_1\right)}^{-1}\right)}\ .
\eal
The coefficient of $z_1^{x}$ in the above generating function  leads to the Skellam distribution. To see this, 
we make use of the summation identity
\bal
e^{\frac{1}{2}z\left(t+t^{-1}\right)}=\sum_{m=-\infty}^{\infty} t^m I_m(z)\,,
\eal
to rewrite the generating function
\bal
G(z_1,\frac{1}{z_1};\l_1,\l_2)&=e^{-(\l_1+\l_2)+\sqrt{\l_1\l_2}\left(\sqrt{\frac{\l_1}{\l_2}} z_1 +{\left(\sqrt{\frac{\l_1}{\l_2}} z_1\right)}^{-1}\right)}\\
&=e^{-(\l_1+\l_2)} \sum_{x=-\infty}^{\infty} \left(\sqrt{\frac{\l_1}{\l_2}} z_1\right)^x I_x(2\sqrt{\l_1\l_2}) .
\eal
The coefficient of $z^x$ is precisely the Skellam distribution~\eqref{sk1}. This gives a clear meaning of the Skellam distribution; it is the distribution of the difference of two separate Poisson random variables. 

With this result, we can proceed to compute the MGF and CF, for this we consider 
\bal
M_x(\F)&=\sum_{-\infty}^{\infty}  P(x,\m_1,\m_2) e^{\F x}\\
&= e^{-(\mu_1+\mu_2)}\sum_{-\infty}^{\infty} 
\left(\frac{\mu_1}{\mu_2}\right)^{x/2}I_{x}(2\sqrt{\mu_1\mu_2}) e^{\F x}\\
&= e^{-(\mu_1+\mu_2)}\sum_{-\infty}^{\infty} 
\left(\frac{\mu_1}{\mu_2}e^{2\F }\right)^{x/2}I_{x}(2\sqrt{\mu_1\mu_2})\\
&=e^{-(\m_1+\m_2)+\sqrt{\m_1\m_2}\left(e^\F\sqrt{\frac{\m_1}{\m_2}}  +{\left(e^{\F}\sqrt{\frac{\m_1}{\m_2}}\right)}^{-1}\right)}\\
& = e^{-(\mu_1+\mu_2)+\mu_1e^\F+\mu_2e^{-\F}}\ .
\eal
The derivation for the CF is similar; one only needs to  substitute $\F\to i \F$.

\section{A meaningless averaging scheme}\label{wrong}

In Euclidean signature, we try to get an effective action from
\bal
e^{-S_{\text{eff}}}&=\sum_{J_1=0}^{\infty}P(J_1,\l_1) e^{-S}\,,
\eal
where
\bal
S=\int d^d x \,\cl\ .
\eal
If we naively consider the measure $P(J_1,\l_1)$ to be the Poisson distribution function of the random variable $J_1$ at each point $x$, the average over the random source is
\bal
\int \cd J_1 \sum_{J_1(x)=0}^{\infty}P(J_1(x),\l_1(x)) e^{ \int d^d x  J_1(x)\f(x) }\ .
\eal
To understand how to perform this average, we first consider discretizing the spacetime so that the above expression reduces to 
\bal
&\left(\prod_{n}\sum_{J_1(x_n)=0}^{\infty}P\left(J_1(x_n),\l_1(x_n)\right)\right) e^{\Delta x^d \sum_n J_1(x_n)\f(x_n) }\\
&=\left(\prod_{n}\sum_{J_1(x_n)=0}^{\infty}P\left(J_1(x_n),\l_1(x_n)\right)\right) \left(\prod_n e^{ J_1(x_n)\f(x_n)\Delta x^d }\right)\ .
\eal
where $n$ collectively denote  the position of a point in the discretized lattice. Here we treat the sources on each individual lattice point to be independent of each other, which is the reason we take a product over the $n$ lattice points. On each point  $x_n$ we sum over all possible values of the sources $J_1(x_n)$. The product is 
\bal
&=\prod_{n}\left(\sum_{J_1(x_n)=0}^{\infty}P\left(J_1(x_n),\l_1(x_n)\right)e^{ J_1(x_n)\f(x_n)\Delta x^d }\right) \\
&=\prod_{n}\left(e^{\l_1(x_n)(e^{\f(x_n)\Delta x^d}-1) }\right)=e^{\sum_{n}\l_1(x_n)(e^{\f(x_n)\Delta x^d}-1) }\,,
\eal
where in the second line we have used the sum over the Poisson distribution for any fixed $x_n$~\eqref{mgps}; this is just the standard expression for the Laplace transform of the Poisson distribution with the parameter  $\l(x_n)$.
However, the result of this direct computation looks strange since the $\Delta^d x$ is further on the shoulder of an extra exponential. This result is difficult to interpret; especially it is not clear how to take it's continuous limit. 

Comparing the derivation in this appendix with the results in section~\ref{randphy}, we find  the crucial difference between the naive random average and the Poisson process is that in the latter case the quantity that obeys the Poisson distribution is the measure on the carrier space, rather than the ``rate of event" $J_1(x)$.  

\section{More about the Poisson process}\label{pps}

\subsection{Random points versus counting measures }

The notion of Poisson process has at least two interpretations. One is similar to the arrival process, like the time at which a bus comes to a bus stop, where it is really considered as a process or a sequence. The other puts more emphasis on a measure theoretical interpretation where the (Poisson) point process is considered as a set of points on the carrier space and the number of points in any given subset naturally leads to a counting measure on the subset. Notice that this counting measure has nothing to do with the original measure of the carrier space; rather it is a characteristic property of the point process only. In this language, the measure $N(dx)$ in~\eqref{countm} is precisely a counting measure, its number counts how many points in the (Poisson) point process appears in the subset volume $dV(x) \subset X$. For this special case it is further a random counting measure in the sense that the number $N(dx)$ is a random function of the subset $dV(x)$. With this proper interpretation, an integral over this counting measure is nothing but~\cite{chiu2013stochastic}
\bal
\int_X N(d x) \f(x)=\sum_{i,x_i\in \Pi
	\cap X}\f(x_i)\,,
\eal     
where $\Pi$ denotes the point process and $x_i$ are thus  point  in the point process that falls in $X$. The LHS emphasizes that the point process $N$ could be considered as a random counting measure, while the RHS emphasizes that the point process can be considered as a counting process whose elements are just a set of discrete points/events.

\subsection{A derivation of the Laplace functional} 

One derivation follows from the property of sums over Poisson processes~\cite{kingman1992poisson}. 
We can consider integrals of the type
\bal
\int_\cm dV(x) J_1(x) f(x) = \int \cj(d x) f(x)\,,
\eal
where the function $f(x)$ takes only a finite number of values $f(x)\equiv f^k(x)=f_1\,,\ldots f_k$. We can then decompose $\cm$ into a union of $A_i$,i.e. $\cm=\bigcup_{i=1}^k A_i$  such that 
\bal
A_i=\{x\in \cm\,, f^k(x)=f_i\}\ .
\eal
Then 
\bal
\int_\cm dV(x) J_1(x) f^k(x) =\sum_{i=1}^{k} \int_{A_i} \cj(d x) f_i=\sum_{i=1}^{k} f_i \int_{A_i} \cj(d x)=\sum_{i=1}^{k} f_i N(A_i)\ .
\eal
Given this, we can compute
\bal
\mathbb{E}\left(e^{u\int_\cm dV(x) J_1(x) f^k(x)}\right)=\mathbb{E}\left(e^{u\sum_{i=1}^{k} f_i N(A_i)}\right)=\prod_{i=1}^k e^{\L(A_i)\left(e^{u f_i}-1\right) }
\eal
where we have used the fact that $N(A_i)$ are mutually independent and each satisfying a Poisson distribution with Poisson parameter $\L(A_i)$. Making use of~\eqref{Ll}, we further rewrite the results into
\bal
\prod_{i=1}^k e^{\int_{A_i}\l(x)dV(x)\left(e^{u f^k(x)}-1\right) }=e^{\int_{\cm}\l(x)dV(x)\left(e^{u f^k(x)}-1\right) }\ .
\eal 
Next we consider a family of this simple functions $f^k$ with increasing numbers, $k$, of values. In fact any continuous function, like the $\f(x)$ in our discussion, can be considered as a limit of such a family of simple functions
\bal
\f(x)=\lim\limits_{k\to \infty} f^k(x)\ .
\eal
Therefore in this limit we have
\bal
\mathbb{E}\left(e^{u\int_\cm dV(x) J_1(x) \f(x)}\right)=e^{\int_{\cm}\l(x)dV(x)\left(e^{u \f(x)}-1\right) }\ .
\eal

\providecommand{\href}[2]{#2}\begingroup\raggedright\endgroup

\end{document}